\documentclass[%
 aip,
 amsmath,amssymb,
 preprint,%
 reprint,%
]{revtex4-1}

\usepackage{graphicx}% Include figure files
\usepackage{dcolumn}% Align table columns on decimal point
\usepackage{bm}% bold math
\usepackage[mathlines]{lineno}% Enable numbering of text and display math
\usepackage[utf8]{inputenc}
\usepackage[T1]{fontenc}
\usepackage{mathptmx}
\usepackage{etoolbox}
\usepackage{xcolor}

\makeatletter
\def\@email#1#2{%
 \endgroup
 \patchcmd{\titleblock@produce}
  {\frontmatter@RRAPformat}
  {\frontmatter@RRAPformat{\produce@RRAP{*#1\href{mailto:#2}{#2}}}\frontmatter@RRAPformat}
  {}{}
}%
\makeatother
\begin{document}

\preprint{AIP/123-QED}

\title{Observation of Chaotic fluctuations in Turbulent Plasma}% Force line breaks with \\

\author{R. Bandyopadhyay}
\email{riddhib@princeton.edu}
\affiliation{Department of Astrophysical Sciences, Princeton University, Princeton, NJ 08544, USA}
 
\author{N.~V. Sarlis}%
\affiliation{Department of Physics, National and Kapodistrian University of Athens, Panepistimiopolis, Zografos, 157 84 Athens, Greece}%

\author{J.~M. Weygand}
\affiliation{Department of Earth, Planetary, and Space Sciences, University of California Los Angeles, Los Angeles, CA 90095-1567, USA}

\author{R.~J. Strangeway}
\affiliation{Department of Earth, Planetary, and Space Sciences, University of California Los Angeles, Los Angeles, CA 90095-1567, USA}

\author{R. B. Torbert}
\affiliation{Physics Department, University of New Hampshire, Durham, New Hampshire 03824, USA}

\author{J.~L. Burch}
\affiliation{Southwest Research Institute, San Antonio, Texas 78238-5166, USA}

\date{\today}% It is always \today, today,

\begin{abstract}
Turbulence is a prevalent phenomenon in space and astrophysical plasmas, often characterized by stochastic fluctuations. While laboratory experiments and numerical simulations have revealed chaotic behavior, in-situ observations of turbulent plasmas in natural environments have predominantly shown highly stochastic signatures. Here, we present unprecedented in-situ evidence of chaotic fluctuations in the turbulent solar wind plasma downstream of the Earth's bow shock. By analyzing the relative location of magnetic-field fluctuations on the permutation entropy-complexity plane (C-H plane), we demonstrate that turbulence in the magnetosheath plasma exhibits characteristics of chaotic fluctuations rather than stochastic behavior, diverging from the expected traits of well-developed turbulence. This finding challenges established notions of plasma turbulence and reveals the need for caution when using the magnetosheath as a laboratory for studying plasma turbulence.
\end{abstract}

\maketitle

%\section{Introduction}\label{sec:intro}
It is often challenging to distinguish time series signals arising from chaotic systems from those generated by stochastic processes. However, it is important to make the distinction to learn about the source of irregularity in the signal. Bandt and Pompe (2002) introduced a method based on ordinal patterns to classify signals into these categories~\citep{Bandt2002PRL_complexity}. Rosso et al. (2007) expanded this approach to construct the complexity-entropy plane (C-H plane), providing a graphical framework for categorizing time series into periodic, chaotic, or stochastic systems~\citep{Rosso2007PRL_noise}.

The C-H method has proven particularly valuable in studying solar wind turbulence, where understanding the source and effects of the dynamics is essential. Previous studies have established solar wind fluctuations as strongly stochastic~\citep{Weck2015PRE_complexity, Weygand2019ApJ_complexity}, supporting its characterization as `well-developed' turbulence, a concept foundational to space plasma turbulence research~\citep{Verscharen2019LRSP}.

The Earth's bow shock arises from the interaction between the Earth's magnetosphere and the interplanetary solar wind flow. The region downstream of the bow shock, known as the magnetosheath, has recently garnered considerable attention for studying plasma turbulence~\citep{Sahraoui2020RMPP_turbulence}. 
Characterized by strong turbulence, often at higher amplitudes compared to the interplanetary solar wind~\citep{Bandyopadhyay2020PRL_hall}, the magnetosheath plasma often exhibit features consistent with fully developed turbulence, including Kolmogorov spectra~\citep{Huang2017ApJL}, constant energy flux through inertial scales~\citep{Hadid2018PRL, Bandyopadhyay2018bApJ}, and strong intermittency~\citep{Chhiber2018JGR}. Several coherent structures such as current sheets~\citep{Chasapis2018ApJ}, Alfv\'en vortices~\citep{Alexandrova2006JGR_Alfven-vortex}, and small-scale magnetic reconnection~\citep{Retino2007NaturePh} are found abundantly.

%\section{Data \& Method}\label{sec:data}
Despite the characterization of magnetosheath plasma as strongly turbulent, our analysis using the permutation entropy-complexity plane (C-H plane) reveals a departure from stochastic behavior towards chaotic fluctuations. 
We compute permutation entropy and Jenson-Shannon statistical complexity from Magnetospheric Multiscale (MMS) mission~\citep{Burch2016SSR, Burch2016Science} data in the magnetosheath region under various conditions. Using burst-mode FIELDS/FGM data~\citep{Russell2016SSR, Torbert2016SSR}, which provide magnetic field measurements at a resolution of $128$ Hz, we analyze $24$ distinct periods selected in the magnetosheath (See APPENDIX A). These periods are selected because they provide relatively long intervals of magnetosheath measurements spanning multiple correlation lengths, and show strong indications of well-developed turbulence. In selecting the data intervals, our goal has been to acquire a reasonably large ensemble of turbulence samples. We eliminate intervals that displayed noticeable coherent wave activity, very low particle number density $n_e < 5 \mathrm{cm}^{-3}$ or very high density  $n_e > 50 \mathrm{cm}^{-3}$. All intervals in this work exhibit an approximate Kolmogorov slope of $-5/3$ in the inertial range of the frequency spectra and a large turbulence amplitude $\delta B / B \gtrsim 1$. However, we note that the frequency band of Kolmogorov slope is narrower than those observed in interplanetary solar wind (See e.g., APPENDIX B).

A particular parameter known to affect the turbulence downstream of the bow shock is the angle made by the interplanetary magnetic field relative to the sunward position of the terrestrial bowshock. The region downstream of the quasi-parallel shock is usually characterized by higher turbulence amplitude compared to those downstream of quasi-perpendicular ones~\citep{Farris1994JGR, shue1998JGR}. Therefore, in this work, we select equal number of intervals of both kinds. Data from each of the four MMS spacecraft are treated independently here. The variation in the complexity and entropy values from each spacecraft’s magnetic field is used as an estimation of the uncertainty. 

The permutation entropy of a given time series is determined using an embedding dimension denoted as $n$. This dimension specifies the size of patterns examined in entropy and complexity calculations. Given a time series of length $T$, the corresponding ordinal pattern probability distribution $p_i$ is established based on all $T-n+1$ length-$n$ segments in the series and all $n!$ permutations of order $n$. The probability of ordinal patterns is computed by counting the number of occurrences of each possible permutation of ordinal patterns in the series
\begin{eqnarray}
    p_i = \frac{\#\{s|\mathrm{s\, is\, the\,} i^{\mathrm{th}}\mathrm{\,ordinal\,pattern}\}}{T-n+1},
\end{eqnarray}
where `$\#$' stands for `the number of'. The permutation entropy $S$ is defined as Shannon’s information entropy for this ordinal pattern probability distribution,
\begin{eqnarray}
S(P) = - \sum_i^{n!} p_i \ln \left( p_i \right),
\end{eqnarray}
where $P$ is the probability distribution function (PDF) of ordinal patterns. The normalized entropy is given by
\begin{eqnarray}
H(P) = \frac{S(P)}{\ln \left( n! \right) }    
\end{eqnarray}
The Jensen-Shannon complexity can be expressed by
\begin{equation}
C(P) =-2 H(P) \frac{ S(\frac{P+P_e}{2})-\frac{1}{2}S(P)-\frac{1}{2}S(P_e)}
{\frac{n! +1}{n!}\ln(n! +1)-2\ln\left[2 (n!)\right]+\ln( n!)},
\end{equation}
where $P_e$ is the maximum permutation entropy state for which every member of the probability distribution has the same value $1/(n!)$.

For a data set with $T$ data points a rough criterion for reliable evaluation of complexity value is given by $T > 5n!$, for a dimension $n$ (See Ref.~\cite{Riedl2013EPJST_entropy}). The time duration of the selected turbulent intervals ranges from 3 to 45 min (Table~\ref{tab:int}). With the MMS-FGM burst mode resolution of 128 Hz this implies at least $\sim 23000$ points in an interval. Therefore, we calculate the permutation entropy and Jensen-Shannon complexity for each interval, using a dimension $n = 5$.

\begin{figure}
\includegraphics[width=\linewidth]{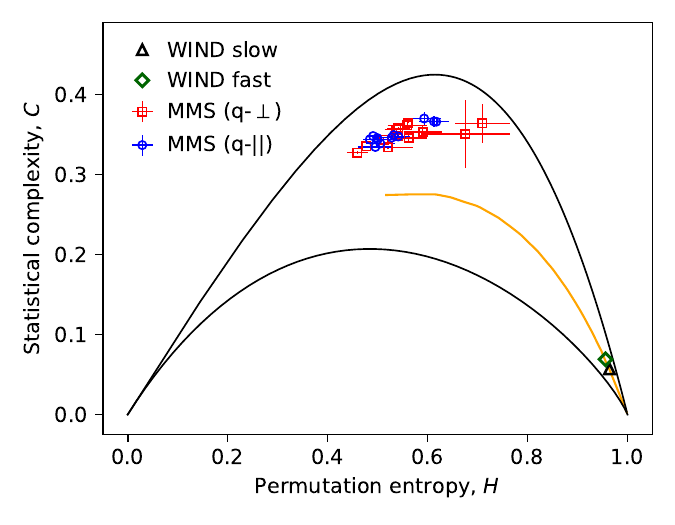}
\caption{\label{fig:ch}The $n = 5$ CH map of turbulent magnetic-field (radial component $B_R$) data sampled in the magnetosheath by MMS. Intervals downstream of quasi-perpendicular and quasi-parallel shocks are represented by red squares and blue circles. Error bars indicate standard deviation of measurements from the four MMS spacecraft. Slow and fast solar wind points, taken from Ref.~\cite{Weck2015PRE_complexity}, are shown by a black triangle and a green diamond marker. Crescent shaped black curves show the maximum and minimum allowed complexity values, and stochastic fractional brownian motion signals are indicated by an orange line.}
\end{figure}

%\section{Results}\label{sec:result}
Figure~\ref{fig:ch} shows the calculated values of complexity and permutation entropy for all the turbulent magnetosheath intervals. Two separate sets of markers in Fig. 1, red squares and blue circles, represent the two different shock geometries. Error bars indicate standard deviations calculated from the  average of 4 MMS spacecraft for each interval. For context, we also include the values obtained for fast and slow stream magnetic fluctuations in the solar wind, taken from Weck \textit{et al.}~\citep{Weck2015PRE_complexity}. 

The relative positions of these measurements show that the magnetosheath fluctuations are characterized by small permutation entropy and high complexity values, far from stochastic-like behavior. All previous works using natural plasmas had placed data points in the lower right corner, far from the MMS estimates~\citep{Weck2015PRE_complexity, Weygand2019ApJ_complexity, Osmane2019JGR_complexity, Good2020ApJ_ICME, Miranda2021ApJ_complexity, Kilpua2022_ICME}. The classification based on quasi-parallel and quasi-perpendicular shocks does not exhibit any clear distinction in the relative positions. The MMS magnetosheath samples used here exhibit powerlaw spectra followed by a steepening in the kinetic range. A high kurtosis value is observed at small scales for all the selected intervals. However, the application of this complexity-entropy plane analysis suggests that the fluctuations may not be fully suitable for studying plasma turbulence, warranting caution.

Our analysis of turbulence in the magnetosheath reveals intriguing insights into the underlying dynamics of plasma fluctuations in this region. The results indicate that the bow shock converts stochastic fluctuations of the pristine solar wind into chaotic behavior downstream. Previous studies have demonstrated that interplanetary shocks reduce the stochasticity of solar wind fluctuations, but do not transition them into the chaotic regime~\citep{Good2020ApJ_ICME}. The terrestrial bow shock may be sufficiently strong to diminish stochasticity to a greater degree, thereby inducing chaotic behavior.

Two key factors may contribute to the observed chaotic behavior. First, the narrow frequency bandwidth of the turbulent magnetic field: a notable characteristic of the magnetosheath turbulence is the small range of inertial-range frequencies. While the interplanetary solar wind exhibits a broader frequency range spanning several decades in the inertial range~\citep{Kiyani2015PTRSA}, magnetosheath spectra typically display a narrower bandwidth, often limited to less than a decade of frequencies. This restricted frequency bandwidth may severely constrain the available phase space of the system, resulting in higher complexity and lower entropy values. The reduced phase space available for plasma fluctuations in the magnetosheath might be responsible for the observed chaotic behavior.

Secondly, the presence of wave modes and coherent structures in the magnetosheath: several kinds of wave modes and coherent structures have been observed in the magnetosheath~\citep{Maruca2018ApJ, Huang2021JGR_anisotropy, Osmane2015GRL_mirror-mode}. The coexistence of various wave modes, although weaker than the background turbulence, may contribute to the observed high complexity and low permutation entropy values. It is plausible that the nonlinear interaction of these wave modes leads to complex, chaotic dynamics within the magnetosheath, adding to the complexity of plasma turbulence in this region. However, coherent structures are not unique to magnetosheath turbulence. The solar wind and other turbulent plasmas naturally generate small-scale coherent structures due to the non-linear processes.

Our study provides novel insights into the dynamics of turbulence in the magnetosheath region and its implications for plasma turbulence research. Through in-situ observations and analysis, we have demonstrated the presence of chaotic fluctuations within turbulent plasmas downstream of Earth's bow shock, challenging conventional notions of plasma turbulence as predominantly stochastic. These results underscore the complex nature of turbulence in space plasmas and its relevance for understanding shock-plasma interactions.

In conclusion, our findings challenge traditional views of magnetosheath turbulence and highlight the need for further research into its underlying dynamics. By characterizing the magnetosheath turbulence as chaotic rather than stochastic, this study opens new avenues for understanding plasma turbulence under different environments. Our research highlights the importance of continued investigation into the complexities of plasma turbulence and its role in shaping the dynamics of space plasmas. By advancing our understanding of these fundamental processes, we can provide new insights into the behavior of astrophysical systems and their interactions with the surrounding environment.\\

%\begin{acknowledgments}
This research was supported in part by the MMS Early Career Award NASA Grant No. 80NSSC21K1458. We are grateful to the MMS instrument teams for cooperation and collaboration in preparing the data. The authors thank the Wind team for the magnetic field~\citep{Lepping1995SSR} and proton moment dataset~\citep{Ogilvie1995SSR}.
%\end{acknowledgments}

\section*{Author Declarations}\label{sec:dec}
\subsection*{Conflict of Interest}
The authors have no conflicts to disclose.

\section*{Data Availability}\label{sec:data}
This study used Level 2 FPI and FIELDS data according to the guidelines set forth by the \textit{MMS} instrumentation team. All data are freely available at 
	\url{https://lasp.colorado.edu/MMS/sdc/}, Refs.~\onlinecite{Burch2016SSR, Russell2016SSR, Pollock2016SSR}. The Wind data, shifted to the Earth’s bow-shock nose, can be found at \url{https://omniweb.gsfc.nasa.gov/}.
 
\section{Appendix A}\label{sec:app_a}
In this appendix, we present the list of intervals used in the main letter. Table~\ref{tab:int} lists the start and end time, along with the corresponding shock type of all the intervals analyzed.

\begin{table}\label{tab:int}
\caption{Magnetosheath time intervals included in the analysis with corresponding shock types.}
\centering
    \begin{ruledtabular}
        \begin{tabular}{c c}
		Time Interval (UTC)  & Shock Type \\
		\hline
		2016-12-11/15:20:14  –  2016-12-11/15:32:34  & quasi-$\parallel$ \\ 
		2017-09-28/06:31:33  –  2017-09-28/07:01:43  & quasi-$\parallel$ \\
		2017-11-10/22:35:43  –  2017-11-10/22:52:03  & quasi-$\parallel$ \\
		2018-02-21/07:48:53  –  2018-02-21/08:01:43  & quasi-$\parallel$ \\
		2018-10-16/14:22:13  –  2018-10-16/14:32:13  & quasi-$\parallel$ \\
		2018-10-28/11:26:33  –  2018-10-28/11:45:23  & quasi-$\parallel$ \\
		2018-12-18/20:36:03  –  2018-12-18/20:45:21  & quasi-$\parallel$ \\
		2019-01-10/13:19:03  –  2019-01-10/13:28:53  & quasi-$\parallel$ \\
		2019-04-05/11:30:53  –  2019-04-05/12:03:12  & quasi-$\parallel$ \\
		2017-01-27/08:02:03  –  2017-01-27/08:08:03  & quasi-$\parallel$ \\
		2017-10-15/22:23:43  –  2017-10-15/22:26:43  & quasi-$\parallel$ \\
		2017-11-10/22:04:43  –  2017-11-10/22:15:33  & quasi-$\parallel$ \\
		2016-01-11/00:57:04  –  2016-01-11/01:00:34  & quasi-$\perp$ \\  
		2016-01-24/23:36:14  –  2016-01-24/23:47:34  & quasi-$\perp$ \\
		2016-12-18/07:08:34  –  2016-12-18/07:18:14  & quasi-$\perp$ \\
		2017-11-22/07:43:43  –  2017-11-22/07:47:23  & quasi-$\perp$ \\
		2017-11-30/17:48:53  –  2017-11-30/17:52:53  & quasi-$\perp$ \\
		2017-12-06/11:06:03  –  2017-12-06/11:12:43  & quasi-$\perp$ \\
		2017-12-18/11:06:23  –  2017-12-18/11:22:03  & quasi-$\perp$ \\
		2018-04-19/05:08:04  –  2018-04-19/05:41:51  & quasi-$\perp$ \\
		2018-10-22/07:23:23  –  2018-10-22/07:35:03  & quasi-$\perp$ \\
		2018-11-01/22:16:13  –  2018-11-01/22:31:43  & quasi-$\perp$ \\
		2018-11-03/04:38:03  –  2018-11-03/04:46:03  & quasi-$\perp$ \\
		2018-11-21/16:10:14  –  2018-11-21/16:55:31  & quasi-$\perp$ \\
\end{tabular}
\end{ruledtabular}
\end{table}

\section{Appendix B}\label{sec:app_b}
\begin{figure}
\includegraphics[width=0.9\linewidth]{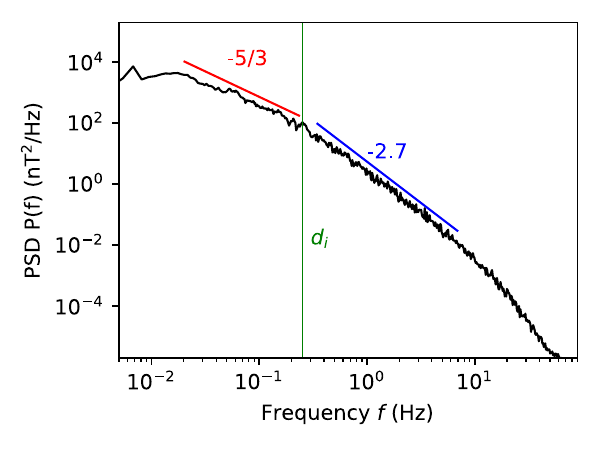}
\caption{\label{fig:psd}{An example of the analyzed power spectral density (PSD) of magnetic fluctuations in the magnetosheath for the time interval 2016-12-11/15:20:14  –  2016-12-11/15:32:34. The vertical green line represent the Taylor-shifted ion inertial length. The Kolmogorov scaling $f^{-5/3}$ and steepening to $f^{-2.7}$ in the sub-ion scales are shown for reference.}}
\end{figure}

Here, we show a sample magnetic-field spectrum for the first interval listed in Table~\ref{tab:int}. Figure~\ref{fig:psd} plots the computed frequency spectrum from MMS/FGM data. A Kolmogorov slope in the inertial range, followed by a steeper slope in the sub-ion scales is clear from the plot.

\section*{References}\label{sec:ref}

%\bibliography{refs_riddhi}% Produces the bibliography via BibTeX.

\begin{thebibliography}{32}%
	\makeatletter
	\providecommand \@ifxundefined [1]{%
		\@ifx{#1\undefined}
	}%
	\providecommand \@ifnum [1]{%
		\ifnum #1\expandafter \@firstoftwo
		\else \expandafter \@secondoftwo
		\fi
	}%
	\providecommand \@ifx [1]{%
		\ifx #1\expandafter \@firstoftwo
		\else \expandafter \@secondoftwo
		\fi
	}%
	\providecommand \natexlab [1]{#1}%
	\providecommand \enquote  [1]{``#1''}%
	\providecommand \bibnamefont  [1]{#1}%
	\providecommand \bibfnamefont [1]{#1}%
	\providecommand \citenamefont [1]{#1}%
	\providecommand \href@noop [0]{\@secondoftwo}%
	\providecommand \href [0]{\begingroup \@sanitize@url \@href}%
	\providecommand \@href[1]{\@@startlink{#1}\@@href}%
	\providecommand \@@href[1]{\endgroup#1\@@endlink}%
	\providecommand \@sanitize@url [0]{\catcode `\\12\catcode `\$12\catcode
		`\&12\catcode `\#12\catcode `\^12\catcode `\_12\catcode `\%12\relax}%
	\providecommand \@@startlink[1]{}%
	\providecommand \@@endlink[0]{}%
	\providecommand \url  [0]{\begingroup\@sanitize@url \@url }%
	\providecommand \@url [1]{\endgroup\@href {#1}{\urlprefix }}%
	\providecommand \urlprefix  [0]{URL }%
	\providecommand \Eprint [0]{\href }%
	\providecommand \doibase [0]{http://dx.doi.org/}%
	\providecommand \selectlanguage [0]{\@gobble}%
	\providecommand \bibinfo  [0]{\@secondoftwo}%
	\providecommand \bibfield  [0]{\@secondoftwo}%
	\providecommand \translation [1]{[#1]}%
	\providecommand \BibitemOpen [0]{}%
	\providecommand \bibitemStop [0]{}%
	\providecommand \bibitemNoStop [0]{.\EOS\space}%
	\providecommand \EOS [0]{\spacefactor3000\relax}%
	\providecommand \BibitemShut  [1]{\csname bibitem#1\endcsname}%
	\let\auto@bib@innerbib\@empty
	%</preamble>
	\bibitem [{\citenamefont {Bandt}\ and\ \citenamefont
		{Pompe}(2002)}]{Bandt2002PRL_complexity}%
	\BibitemOpen
	\bibfield  {author} {\bibinfo {author} {\bibfnamefont {C.}~\bibnamefont
			{Bandt}}\ and\ \bibinfo {author} {\bibfnamefont {B.}~\bibnamefont {Pompe}},\
	}\bibfield  {title} {\enquote {\bibinfo {title} {Permutation entropy: A
				natural complexity measure for time series},}\ }\href {\doibase
		10.1103/PhysRevLett.88.174102} {\bibfield  {journal} {\bibinfo  {journal}
			{Phys. Rev. Lett.}\ }\textbf {\bibinfo {volume} {88}},\ \bibinfo {pages}
		{174102} (\bibinfo {year} {2002})}\BibitemShut {NoStop}%
	\bibitem [{\citenamefont {Rosso}\ \emph {et~al.}(2007)\citenamefont {Rosso},
		\citenamefont {Larrondo}, \citenamefont {Martin}, \citenamefont {Plastino},\
		and\ \citenamefont {Fuentes}}]{Rosso2007PRL_noise}%
	\BibitemOpen
	\bibfield  {author} {\bibinfo {author} {\bibfnamefont {O.~A.}\ \bibnamefont
			{Rosso}}, \bibinfo {author} {\bibfnamefont {H.~A.}\ \bibnamefont {Larrondo}},
		\bibinfo {author} {\bibfnamefont {M.~T.}\ \bibnamefont {Martin}}, \bibinfo
		{author} {\bibfnamefont {A.}~\bibnamefont {Plastino}}, \ and\ \bibinfo
		{author} {\bibfnamefont {M.~A.}\ \bibnamefont {Fuentes}},\ }\bibfield
	{title} {\enquote {\bibinfo {title} {Distinguishing noise from chaos},}\
	}\href {\doibase 10.1103/PhysRevLett.99.154102} {\bibfield  {journal}
		{\bibinfo  {journal} {Phys. Rev. Lett.}\ }\textbf {\bibinfo {volume} {99}},\
		\bibinfo {pages} {154102} (\bibinfo {year} {2007})}\BibitemShut {NoStop}%
	\bibitem [{\citenamefont {Weck}\ \emph {et~al.}(2015)\citenamefont {Weck},
		\citenamefont {Schaffner}, \citenamefont {Brown},\ and\ \citenamefont
		{Wicks}}]{Weck2015PRE_complexity}%
	\BibitemOpen
	\bibfield  {author} {\bibinfo {author} {\bibfnamefont {P.~J.}\ \bibnamefont
			{Weck}}, \bibinfo {author} {\bibfnamefont {D.~A.}\ \bibnamefont {Schaffner}},
		\bibinfo {author} {\bibfnamefont {M.~R.}\ \bibnamefont {Brown}}, \ and\
		\bibinfo {author} {\bibfnamefont {R.~T.}\ \bibnamefont {Wicks}},\ }\bibfield
	{title} {\enquote {\bibinfo {title} {Permutation entropy and statistical
				complexity analysis of turbulence in laboratory plasmas and the solar
				wind},}\ }\href {\doibase 10.1103/PhysRevE.91.023101} {\bibfield  {journal}
		{\bibinfo  {journal} {Phys. Rev. E}\ }\textbf {\bibinfo {volume} {91}},\
		\bibinfo {pages} {023101} (\bibinfo {year} {2015})}\BibitemShut {NoStop}%
	\bibitem [{\citenamefont {Weygand}\ and\ \citenamefont
		{Kivelson}(2019)}]{Weygand2019ApJ_complexity}%
	\BibitemOpen
	\bibfield  {author} {\bibinfo {author} {\bibfnamefont {J.~M.}\ \bibnamefont
			{Weygand}}\ and\ \bibinfo {author} {\bibfnamefont {M.~G.}\ \bibnamefont
			{Kivelson}},\ }\bibfield  {title} {\enquote {\bibinfo {title}
			{Jensen–shannon complexity measurements in solar wind magnetic field
				fluctuations},}\ }\href {\doibase 10.3847/1538-4357/aafda4} {\bibfield
		{journal} {\bibinfo  {journal} {The Astrophysical Journal}\ }\textbf
		{\bibinfo {volume} {872}},\ \bibinfo {pages} {59} (\bibinfo {year}
		{2019})}\BibitemShut {NoStop}%
	\bibitem [{\citenamefont {{Verscharen}}, \citenamefont {{Klein}},\ and\
		\citenamefont {{Maruca}}(2019)}]{Verscharen2019LRSP}%
	\BibitemOpen
	\bibfield  {author} {\bibinfo {author} {\bibfnamefont {D.}~\bibnamefont
			{{Verscharen}}}, \bibinfo {author} {\bibfnamefont {K.~G.}\ \bibnamefont
			{{Klein}}}, \ and\ \bibinfo {author} {\bibfnamefont {B.~A.}\ \bibnamefont
			{{Maruca}}},\ }\bibfield  {title} {\enquote {\bibinfo {title} {The
				multi-scale nature of the solar wind},}\ }\href {\doibase
		10.1007/s41116-019-0021-0} {\bibfield  {journal} {\bibinfo  {journal} {Living
				Reviews in Solar Physics}\ }\textbf {\bibinfo {volume} {16}},\ \bibinfo
		{pages} {5} (\bibinfo {year} {2019})}\BibitemShut {NoStop}%
	\bibitem [{\citenamefont {{Sahraoui}}, \citenamefont {{Hadid}},\ and\
		\citenamefont {{Huang}}(2020)}]{Sahraoui2020RMPP_turbulence}%
	\BibitemOpen
	\bibfield  {author} {\bibinfo {author} {\bibfnamefont {F.}~\bibnamefont
			{{Sahraoui}}}, \bibinfo {author} {\bibfnamefont {L.}~\bibnamefont {{Hadid}}},
		\ and\ \bibinfo {author} {\bibfnamefont {S.}~\bibnamefont {{Huang}}},\
	}\bibfield  {title} {\enquote {\bibinfo {title} {{Magnetohydrodynamic and
					kinetic scale turbulence in the near-Earth space plasmas: a (short) biased
					review}},}\ }\href {\doibase 10.1007/s41614-020-0040-2} {\bibfield  {journal}
		{\bibinfo  {journal} {Reviews of Modern Plasma Physics}\ }\textbf {\bibinfo
			{volume} {4}},\ \bibinfo {eid} {4} (\bibinfo {year} {2020})}\BibitemShut
	{NoStop}%
	\bibitem [{\citenamefont {Bandyopadhyay}\ \emph {et~al.}(2020)\citenamefont
		{Bandyopadhyay}, \citenamefont {Sorriso-Valvo}, \citenamefont {Chasapis},
		\citenamefont {Hellinger}, \citenamefont {Matthaeus}, \citenamefont
		{Verdini}, \citenamefont {Landi}, \citenamefont {Franci}, \citenamefont
		{Matteini}, \citenamefont {Giles}, \citenamefont {Gershman}, \citenamefont
		{Moore}, \citenamefont {Pollock}, \citenamefont {Russell}, \citenamefont
		{Strangeway}, \citenamefont {Torbert},\ and\ \citenamefont
		{Burch}}]{Bandyopadhyay2020PRL_hall}%
	\BibitemOpen
	\bibfield  {author} {\bibinfo {author} {\bibfnamefont {R.}~\bibnamefont
			{Bandyopadhyay}}, \bibinfo {author} {\bibfnamefont {L.}~\bibnamefont
			{Sorriso-Valvo}}, \bibinfo {author} {\bibfnamefont {A.}~\bibnamefont
			{Chasapis}}, \bibinfo {author} {\bibfnamefont {P.}~\bibnamefont {Hellinger}},
		\bibinfo {author} {\bibfnamefont {W.~H.}\ \bibnamefont {Matthaeus}}, \bibinfo
		{author} {\bibfnamefont {A.}~\bibnamefont {Verdini}}, \bibinfo {author}
		{\bibfnamefont {S.}~\bibnamefont {Landi}}, \bibinfo {author} {\bibfnamefont
			{L.}~\bibnamefont {Franci}}, \bibinfo {author} {\bibfnamefont
			{L.}~\bibnamefont {Matteini}}, \bibinfo {author} {\bibfnamefont {B.~L.}\
			\bibnamefont {Giles}}, \bibinfo {author} {\bibfnamefont {D.~J.}\ \bibnamefont
			{Gershman}}, \bibinfo {author} {\bibfnamefont {T.~E.}\ \bibnamefont {Moore}},
		\bibinfo {author} {\bibfnamefont {C.~J.}\ \bibnamefont {Pollock}}, \bibinfo
		{author} {\bibfnamefont {C.~T.}\ \bibnamefont {Russell}}, \bibinfo {author}
		{\bibfnamefont {R.~J.}\ \bibnamefont {Strangeway}}, \bibinfo {author}
		{\bibfnamefont {R.~B.}\ \bibnamefont {Torbert}}, \ and\ \bibinfo {author}
		{\bibfnamefont {J.~L.}\ \bibnamefont {Burch}},\ }\bibfield  {title} {\enquote
		{\bibinfo {title} {{In Situ Observation of Hall Magnetohydrodynamic Cascade
					in Space Plasma}},}\ }\href {\doibase 10.1103/PhysRevLett.124.225101}
	{\bibfield  {journal} {\bibinfo  {journal} {Physical Review Letters}\
		}\textbf {\bibinfo {volume} {124}},\ \bibinfo {pages} {225101} (\bibinfo
		{year} {2020})}\BibitemShut {NoStop}%
	\bibitem [{\citenamefont {Huang}\ \emph {et~al.}(2017)\citenamefont {Huang},
		\citenamefont {Hadid}, \citenamefont {Sahraoui}, \citenamefont {Yuan},\ and\
		\citenamefont {Deng}}]{Huang2017ApJL}%
	\BibitemOpen
	\bibfield  {author} {\bibinfo {author} {\bibfnamefont {S.~Y.}\ \bibnamefont
			{Huang}}, \bibinfo {author} {\bibfnamefont {L.~Z.}\ \bibnamefont {Hadid}},
		\bibinfo {author} {\bibfnamefont {F.}~\bibnamefont {Sahraoui}}, \bibinfo
		{author} {\bibfnamefont {Z.~G.}\ \bibnamefont {Yuan}}, \ and\ \bibinfo
		{author} {\bibfnamefont {X.~H.}\ \bibnamefont {Deng}},\ }\bibfield  {title}
	{\enquote {\bibinfo {title} {On the existence of the kolmogorov inertial
				range in the terrestrial magnetosheath turbulence},}\ }\href
	{http://stacks.iop.org/2041-8205/836/i=1/a=L10} {\bibfield  {journal}
		{\bibinfo  {journal} {Astrophys. J. Lett.}\ }\textbf {\bibinfo {volume}
			{836}},\ \bibinfo {pages} {L10} (\bibinfo {year} {2017})}\BibitemShut
	{NoStop}%
	\bibitem [{\citenamefont {Hadid}\ \emph {et~al.}(2018)\citenamefont {Hadid},
		\citenamefont {Sahraoui}, \citenamefont {Galtier},\ and\ \citenamefont
		{Huang}}]{Hadid2018PRL}%
	\BibitemOpen
	\bibfield  {author} {\bibinfo {author} {\bibfnamefont {L.~Z.}\ \bibnamefont
			{Hadid}}, \bibinfo {author} {\bibfnamefont {F.}~\bibnamefont {Sahraoui}},
		\bibinfo {author} {\bibfnamefont {S.}~\bibnamefont {Galtier}}, \ and\
		\bibinfo {author} {\bibfnamefont {S.~Y.}\ \bibnamefont {Huang}},\ }\bibfield
	{title} {\enquote {\bibinfo {title} {Compressible magnetohydrodynamic
				turbulence in the earth's magnetosheath: Estimation of the energy cascade
				rate using in situ spacecraft data},}\ }\href {\doibase
		10.1103/PhysRevLett.120.055102} {\bibfield  {journal} {\bibinfo  {journal}
			{Phys. Rev. Lett.}\ }\textbf {\bibinfo {volume} {120}},\ \bibinfo {pages}
		{055102} (\bibinfo {year} {2018})}\BibitemShut {NoStop}%
	\bibitem [{\citenamefont {Bandyopadhyay}\ \emph {et~al.}(2018)\citenamefont
		{Bandyopadhyay}, \citenamefont {Chasapis}, \citenamefont {Chhiber},
		\citenamefont {Parashar}, \citenamefont {Matthaeus}, \citenamefont {Shay},
		\citenamefont {Maruca}, \citenamefont {Burch}, \citenamefont {Moore},
		\citenamefont {Pollock}, \citenamefont {Giles}, \citenamefont {Paterson},
		\citenamefont {Dorelli}, \citenamefont {Gershman}, \citenamefont {Torbert},
		\citenamefont {Russell},\ and\ \citenamefont
		{Strangeway}}]{Bandyopadhyay2018bApJ}%
	\BibitemOpen
	\bibfield  {author} {\bibinfo {author} {\bibfnamefont {R.}~\bibnamefont
			{Bandyopadhyay}}, \bibinfo {author} {\bibfnamefont {A.}~\bibnamefont
			{Chasapis}}, \bibinfo {author} {\bibfnamefont {R.}~\bibnamefont {Chhiber}},
		\bibinfo {author} {\bibfnamefont {T.~N.}\ \bibnamefont {Parashar}}, \bibinfo
		{author} {\bibfnamefont {W.~H.}\ \bibnamefont {Matthaeus}}, \bibinfo {author}
		{\bibfnamefont {M.~A.}\ \bibnamefont {Shay}}, \bibinfo {author}
		{\bibfnamefont {B.~A.}\ \bibnamefont {Maruca}}, \bibinfo {author}
		{\bibfnamefont {J.~L.}\ \bibnamefont {Burch}}, \bibinfo {author}
		{\bibfnamefont {T.~E.}\ \bibnamefont {Moore}}, \bibinfo {author}
		{\bibfnamefont {C.~J.}\ \bibnamefont {Pollock}}, \bibinfo {author}
		{\bibfnamefont {B.~L.}\ \bibnamefont {Giles}}, \bibinfo {author}
		{\bibfnamefont {W.~R.}\ \bibnamefont {Paterson}}, \bibinfo {author}
		{\bibfnamefont {J.}~\bibnamefont {Dorelli}}, \bibinfo {author} {\bibfnamefont
			{D.~J.}\ \bibnamefont {Gershman}}, \bibinfo {author} {\bibfnamefont {R.~B.}\
			\bibnamefont {Torbert}}, \bibinfo {author} {\bibfnamefont {C.~T.}\
			\bibnamefont {Russell}}, \ and\ \bibinfo {author} {\bibfnamefont {R.~J.}\
			\bibnamefont {Strangeway}},\ }\bibfield  {title} {\enquote {\bibinfo {title}
			{{Incompressive Energy Transfer in the Earth's Magnetosheath: Magnetospheric
					Multiscale Observations}},}\ }\href {\doibase 10.3847/1538-4357/aade04}
	{\bibfield  {journal} {\bibinfo  {journal} {The Astrophysical Journal}\
		}\textbf {\bibinfo {volume} {866}},\ \bibinfo {pages} {106} (\bibinfo {year}
		{2018})}\BibitemShut {NoStop}%
	\bibitem [{\citenamefont {Chhiber}\ \emph {et~al.}(2018)\citenamefont
		{Chhiber}, \citenamefont {Chasapis}, \citenamefont {Bandyopadhyay},
		\citenamefont {Parashar}, \citenamefont {Matthaeus}, \citenamefont {Maruca},
		\citenamefont {Moore}, \citenamefont {Burch}, \citenamefont {Torbert},
		\citenamefont {Russell}, \citenamefont {Le~Contel}, \citenamefont {Argall},
		\citenamefont {Fischer}, \citenamefont {Mirioni}, \citenamefont {Strangeway},
		\citenamefont {Pollock}, \citenamefont {Giles},\ and\ \citenamefont
		{Gershman}}]{Chhiber2018JGR}%
	\BibitemOpen
	\bibfield  {author} {\bibinfo {author} {\bibfnamefont {R.}~\bibnamefont
			{Chhiber}}, \bibinfo {author} {\bibfnamefont {A.}~\bibnamefont {Chasapis}},
		\bibinfo {author} {\bibfnamefont {R.}~\bibnamefont {Bandyopadhyay}}, \bibinfo
		{author} {\bibfnamefont {T.~N.}\ \bibnamefont {Parashar}}, \bibinfo {author}
		{\bibfnamefont {W.~H.}\ \bibnamefont {Matthaeus}}, \bibinfo {author}
		{\bibfnamefont {B.~A.}\ \bibnamefont {Maruca}}, \bibinfo {author}
		{\bibfnamefont {T.~E.}\ \bibnamefont {Moore}}, \bibinfo {author}
		{\bibfnamefont {J.~L.}\ \bibnamefont {Burch}}, \bibinfo {author}
		{\bibfnamefont {R.~B.}\ \bibnamefont {Torbert}}, \bibinfo {author}
		{\bibfnamefont {C.~T.}\ \bibnamefont {Russell}}, \bibinfo {author}
		{\bibfnamefont {O.}~\bibnamefont {Le~Contel}}, \bibinfo {author}
		{\bibfnamefont {M.~R.}\ \bibnamefont {Argall}}, \bibinfo {author}
		{\bibfnamefont {D.}~\bibnamefont {Fischer}}, \bibinfo {author} {\bibfnamefont
			{L.}~\bibnamefont {Mirioni}}, \bibinfo {author} {\bibfnamefont {R.~J.}\
			\bibnamefont {Strangeway}}, \bibinfo {author} {\bibfnamefont {C.~J.}\
			\bibnamefont {Pollock}}, \bibinfo {author} {\bibfnamefont {B.~L.}\
			\bibnamefont {Giles}}, \ and\ \bibinfo {author} {\bibfnamefont {D.~J.}\
			\bibnamefont {Gershman}},\ }\bibfield  {title} {\enquote {\bibinfo {title}
			{Higher-order turbulence statistics in the earth's magnetosheath and the
				solar wind using magnetospheric multiscale observations},}\ }\href {\doibase
		10.1029/2018JA025768} {\bibfield  {journal} {\bibinfo  {journal} {Journal of
				Geophysical Research: Space Physics}\ }\textbf {\bibinfo {volume} {123}},\
		\bibinfo {pages} {9941--9954} (\bibinfo {year} {2018})}\BibitemShut {NoStop}%
	\bibitem [{\citenamefont {Chasapis}\ \emph {et~al.}(2018)\citenamefont
		{Chasapis}, \citenamefont {Yang}, \citenamefont {Matthaeus}, \citenamefont
		{Parashar}, \citenamefont {Haggerty}, \citenamefont {Burch}, \citenamefont
		{Moore}, \citenamefont {Pollock}, \citenamefont {Dorelli}, \citenamefont
		{Gershman}, \citenamefont {Torbert},\ and\ \citenamefont
		{Russell}}]{Chasapis2018ApJ}%
	\BibitemOpen
	\bibfield  {author} {\bibinfo {author} {\bibfnamefont {A.}~\bibnamefont
			{Chasapis}}, \bibinfo {author} {\bibfnamefont {Y.}~\bibnamefont {Yang}},
		\bibinfo {author} {\bibfnamefont {W.~H.}\ \bibnamefont {Matthaeus}}, \bibinfo
		{author} {\bibfnamefont {T.~N.}\ \bibnamefont {Parashar}}, \bibinfo {author}
		{\bibfnamefont {C.~C.}\ \bibnamefont {Haggerty}}, \bibinfo {author}
		{\bibfnamefont {J.~L.}\ \bibnamefont {Burch}}, \bibinfo {author}
		{\bibfnamefont {T.~E.}\ \bibnamefont {Moore}}, \bibinfo {author}
		{\bibfnamefont {C.~J.}\ \bibnamefont {Pollock}}, \bibinfo {author}
		{\bibfnamefont {J.}~\bibnamefont {Dorelli}}, \bibinfo {author} {\bibfnamefont
			{D.~J.}\ \bibnamefont {Gershman}}, \bibinfo {author} {\bibfnamefont {R.~B.}\
			\bibnamefont {Torbert}}, \ and\ \bibinfo {author} {\bibfnamefont {C.~T.}\
			\bibnamefont {Russell}},\ }\bibfield  {title} {\enquote {\bibinfo {title}
			{Energy conversion and collisionless plasma dissipation channels in the
				turbulent magnetosheath observed by the magnetospheric multiscale mission},}\
	}\href {\doibase 10.3847/1538-4357/aac775} {\bibfield  {journal} {\bibinfo
			{journal} {The Astrophysical Journal}\ }\textbf {\bibinfo {volume} {862}},\
		\bibinfo {pages} {32} (\bibinfo {year} {2018})}\BibitemShut {NoStop}%
	\bibitem [{\citenamefont {{Alexandrova}}\ \emph {et~al.}(2006)\citenamefont
		{{Alexandrova}}, \citenamefont {{Mangeney}}, \citenamefont {{Maksimovic}},
		\citenamefont {{Cornilleau-Wehrlin}}, \citenamefont {{Bosqued}},\ and\
		\citenamefont {{Andr{\'e}}}}]{Alexandrova2006JGR_Alfven-vortex}%
	\BibitemOpen
	\bibfield  {author} {\bibinfo {author} {\bibfnamefont {O.}~\bibnamefont
			{{Alexandrova}}}, \bibinfo {author} {\bibfnamefont {A.}~\bibnamefont
			{{Mangeney}}}, \bibinfo {author} {\bibfnamefont {M.}~\bibnamefont
			{{Maksimovic}}}, \bibinfo {author} {\bibfnamefont {N.}~\bibnamefont
			{{Cornilleau-Wehrlin}}}, \bibinfo {author} {\bibfnamefont {J.~M.}\
			\bibnamefont {{Bosqued}}}, \ and\ \bibinfo {author} {\bibfnamefont
			{M.}~\bibnamefont {{Andr{\'e}}}},\ }\bibfield  {title} {\enquote {\bibinfo
			{title} {{Alfv{\'e}n vortex filaments observed in magnetosheath downstream of
					a quasi-perpendicular bow shock}},}\ }\href {\doibase 10.1029/2006JA011934}
	{\bibfield  {journal} {\bibinfo  {journal} {Journal of Geophysical Research
				(Space Physics)}\ }\textbf {\bibinfo {volume} {111}},\ \bibinfo {eid}
		{A12208} (\bibinfo {year} {2006})}\BibitemShut {NoStop}%
	\bibitem [{\citenamefont {{Retin{\`o}}}\ \emph {et~al.}(2007)\citenamefont
		{{Retin{\`o}}}, \citenamefont {{Sundkvist}}, \citenamefont {{Vaivads}},
		\citenamefont {{Mozer}}, \citenamefont {{Andr{\'e}}},\ and\ \citenamefont
		{{Owen}}}]{Retino2007NaturePh}%
	\BibitemOpen
	\bibfield  {author} {\bibinfo {author} {\bibfnamefont {A.}~\bibnamefont
			{{Retin{\`o}}}}, \bibinfo {author} {\bibfnamefont {D.}~\bibnamefont
			{{Sundkvist}}}, \bibinfo {author} {\bibfnamefont {A.}~\bibnamefont
			{{Vaivads}}}, \bibinfo {author} {\bibfnamefont {F.}~\bibnamefont {{Mozer}}},
		\bibinfo {author} {\bibfnamefont {M.}~\bibnamefont {{Andr{\'e}}}}, \ and\
		\bibinfo {author} {\bibfnamefont {C.~J.}\ \bibnamefont {{Owen}}},\ }\bibfield
	{title} {\enquote {\bibinfo {title} {{In situ evidence of magnetic
					reconnection in turbulent plasma}},}\ }\href {\doibase 10.1038/nphys574}
	{\bibfield  {journal} {\bibinfo  {journal} {Nature Physics}\ }\textbf
		{\bibinfo {volume} {3}},\ \bibinfo {pages} {236--238} (\bibinfo {year}
		{2007})}\BibitemShut {NoStop}%
	\bibitem [{\citenamefont {Burch}\ \emph {et~al.}(2016)\citenamefont {Burch},
		\citenamefont {Moore}, \citenamefont {Torbert},\ and\ \citenamefont
		{Giles}}]{Burch2016SSR}%
	\BibitemOpen
	\bibfield  {author} {\bibinfo {author} {\bibfnamefont {J.~L.}\ \bibnamefont
			{Burch}}, \bibinfo {author} {\bibfnamefont {T.~E.}\ \bibnamefont {Moore}},
		\bibinfo {author} {\bibfnamefont {R.~B.}\ \bibnamefont {Torbert}}, \ and\
		\bibinfo {author} {\bibfnamefont {B.~L.}\ \bibnamefont {Giles}},\ }\bibfield
	{title} {\enquote {\bibinfo {title} {Magnetospheric multiscale overview and
				science objectives},}\ }\href {\doibase 10.1007/s11214-015-0164-9} {\bibfield
		{journal} {\bibinfo  {journal} {Space Science Reviews}\ }\textbf {\bibinfo
			{volume} {199}},\ \bibinfo {pages} {5--21} (\bibinfo {year}
		{2016})}\BibitemShut {NoStop}%
	\bibitem [{\citenamefont {{Burch}}\ \emph {et~al.}(2016)\citenamefont
		{{Burch}}, \citenamefont {{Torbert}}, \citenamefont {{Phan}}, \citenamefont
		{{Chen}}, \citenamefont {{Moore}}, \citenamefont {{Ergun}}, \citenamefont
		{{Eastwood}}, \citenamefont {{Gershman}}, \citenamefont {{Cassak}},
		\citenamefont {{Argall}}, \citenamefont {{Wang}}, \citenamefont {{Hesse}},
		\citenamefont {{Pollock}}, \citenamefont {{Giles}}, \citenamefont
		{{Nakamura}}, \citenamefont {{Mauk}}, \citenamefont {{Fuselier}},
		\citenamefont {{Russell}}, \citenamefont {{Strangeway}}, \citenamefont
		{{Drake}}, \citenamefont {{Shay}}, \citenamefont {{Khotyaintsev}},
		\citenamefont {{Lindqvist}}, \citenamefont {{Marklund}}, \citenamefont
		{{Wilder}}, \citenamefont {{Young}}, \citenamefont {{Torkar}}, \citenamefont
		{{Goldstein}}, \citenamefont {{Dorelli}}, \citenamefont {{Avanov}},
		\citenamefont {{Oka}}, \citenamefont {{Baker}}, \citenamefont {{Jaynes}},
		\citenamefont {{Goodrich}}, \citenamefont {{Cohen}}, \citenamefont
		{{Turner}}, \citenamefont {{Fennell}}, \citenamefont {{Blake}}, \citenamefont
		{{Clemmons}}, \citenamefont {{Goldman}}, \citenamefont {{Newman}},
		\citenamefont {{Petrinec}}, \citenamefont {{Trattner}}, \citenamefont
		{{Lavraud}}, \citenamefont {{Reiff}}, \citenamefont {{Baumjohann}},
		\citenamefont {{Magnes}}, \citenamefont {{Steller}}, \citenamefont {{Lewis}},
		\citenamefont {{Saito}}, \citenamefont {{Coffey}},\ and\ \citenamefont
		{{Chandler}}}]{Burch2016Science}%
	\BibitemOpen
	\bibfield  {author} {\bibinfo {author} {\bibfnamefont {J.~L.}\ \bibnamefont
			{{Burch}}}, \bibinfo {author} {\bibfnamefont {R.~B.}\ \bibnamefont
			{{Torbert}}}, \bibinfo {author} {\bibfnamefont {T.~D.}\ \bibnamefont
			{{Phan}}}, \bibinfo {author} {\bibfnamefont {L.~J.}\ \bibnamefont {{Chen}}},
		\bibinfo {author} {\bibfnamefont {T.~E.}\ \bibnamefont {{Moore}}}, \bibinfo
		{author} {\bibfnamefont {R.~E.}\ \bibnamefont {{Ergun}}}, \bibinfo {author}
		{\bibfnamefont {J.~P.}\ \bibnamefont {{Eastwood}}}, \bibinfo {author}
		{\bibfnamefont {D.~J.}\ \bibnamefont {{Gershman}}}, \bibinfo {author}
		{\bibfnamefont {P.~A.}\ \bibnamefont {{Cassak}}}, \bibinfo {author}
		{\bibfnamefont {M.~R.}\ \bibnamefont {{Argall}}}, \bibinfo {author}
		{\bibfnamefont {S.}~\bibnamefont {{Wang}}}, \bibinfo {author} {\bibfnamefont
			{M.}~\bibnamefont {{Hesse}}}, \bibinfo {author} {\bibfnamefont {C.~J.}\
			\bibnamefont {{Pollock}}}, \bibinfo {author} {\bibfnamefont {B.~L.}\
			\bibnamefont {{Giles}}}, \bibinfo {author} {\bibfnamefont {R.}~\bibnamefont
			{{Nakamura}}}, \bibinfo {author} {\bibfnamefont {B.~H.}\ \bibnamefont
			{{Mauk}}}, \bibinfo {author} {\bibfnamefont {S.~A.}\ \bibnamefont
			{{Fuselier}}}, \bibinfo {author} {\bibfnamefont {C.~T.}\ \bibnamefont
			{{Russell}}}, \bibinfo {author} {\bibfnamefont {R.~J.}\ \bibnamefont
			{{Strangeway}}}, \bibinfo {author} {\bibfnamefont {J.~F.}\ \bibnamefont
			{{Drake}}}, \bibinfo {author} {\bibfnamefont {M.~A.}\ \bibnamefont {{Shay}}},
		\bibinfo {author} {\bibfnamefont {Y.~V.}\ \bibnamefont {{Khotyaintsev}}},
		\bibinfo {author} {\bibfnamefont {P.~A.}\ \bibnamefont {{Lindqvist}}},
		\bibinfo {author} {\bibfnamefont {G.}~\bibnamefont {{Marklund}}}, \bibinfo
		{author} {\bibfnamefont {F.~D.}\ \bibnamefont {{Wilder}}}, \bibinfo {author}
		{\bibfnamefont {D.~T.}\ \bibnamefont {{Young}}}, \bibinfo {author}
		{\bibfnamefont {K.}~\bibnamefont {{Torkar}}}, \bibinfo {author}
		{\bibfnamefont {J.}~\bibnamefont {{Goldstein}}}, \bibinfo {author}
		{\bibfnamefont {J.~C.}\ \bibnamefont {{Dorelli}}}, \bibinfo {author}
		{\bibfnamefont {L.~A.}\ \bibnamefont {{Avanov}}}, \bibinfo {author}
		{\bibfnamefont {M.}~\bibnamefont {{Oka}}}, \bibinfo {author} {\bibfnamefont
			{D.~N.}\ \bibnamefont {{Baker}}}, \bibinfo {author} {\bibfnamefont {A.~N.}\
			\bibnamefont {{Jaynes}}}, \bibinfo {author} {\bibfnamefont {K.~A.}\
			\bibnamefont {{Goodrich}}}, \bibinfo {author} {\bibfnamefont {I.~J.}\
			\bibnamefont {{Cohen}}}, \bibinfo {author} {\bibfnamefont {D.~L.}\
			\bibnamefont {{Turner}}}, \bibinfo {author} {\bibfnamefont {J.~F.}\
			\bibnamefont {{Fennell}}}, \bibinfo {author} {\bibfnamefont {J.~B.}\
			\bibnamefont {{Blake}}}, \bibinfo {author} {\bibfnamefont {J.}~\bibnamefont
			{{Clemmons}}}, \bibinfo {author} {\bibfnamefont {M.}~\bibnamefont
			{{Goldman}}}, \bibinfo {author} {\bibfnamefont {D.}~\bibnamefont {{Newman}}},
		\bibinfo {author} {\bibfnamefont {S.~M.}\ \bibnamefont {{Petrinec}}},
		\bibinfo {author} {\bibfnamefont {K.~J.}\ \bibnamefont {{Trattner}}},
		\bibinfo {author} {\bibfnamefont {B.}~\bibnamefont {{Lavraud}}}, \bibinfo
		{author} {\bibfnamefont {P.~H.}\ \bibnamefont {{Reiff}}}, \bibinfo {author}
		{\bibfnamefont {W.}~\bibnamefont {{Baumjohann}}}, \bibinfo {author}
		{\bibfnamefont {W.}~\bibnamefont {{Magnes}}}, \bibinfo {author}
		{\bibfnamefont {M.}~\bibnamefont {{Steller}}}, \bibinfo {author}
		{\bibfnamefont {W.}~\bibnamefont {{Lewis}}}, \bibinfo {author} {\bibfnamefont
			{Y.}~\bibnamefont {{Saito}}}, \bibinfo {author} {\bibfnamefont
			{V.}~\bibnamefont {{Coffey}}}, \ and\ \bibinfo {author} {\bibfnamefont
			{M.}~\bibnamefont {{Chandler}}},\ }\bibfield  {title} {\enquote {\bibinfo
			{title} {Electron-scale measurements of magnetic reconnection in space},}\
	}\href {\doibase 10.1126/science.aaf2939} {\bibfield  {journal} {\bibinfo
			{journal} {Science}\ }\textbf {\bibinfo {volume} {352}},\ \bibinfo {eid}
		{aaf2939} (\bibinfo {year} {2016})}\BibitemShut {NoStop}%
	\bibitem [{\citenamefont {Russell}\ \emph {et~al.}(2016)\citenamefont
		{Russell}, \citenamefont {Anderson}, \citenamefont {Baumjohann},
		\citenamefont {Bromund}, \citenamefont {Dearborn}, \citenamefont {Fischer},
		\citenamefont {Le}, \citenamefont {Leinweber}, \citenamefont {Leneman},
		\citenamefont {Magnes}, \citenamefont {Means}, \citenamefont {Moldwin},
		\citenamefont {Nakamura}, \citenamefont {Pierce}, \citenamefont {Plaschke},
		\citenamefont {Rowe}, \citenamefont {Slavin}, \citenamefont {Strangeway},
		\citenamefont {Torbert}, \citenamefont {Hagen}, \citenamefont {Jernej},
		\citenamefont {Valavanoglou},\ and\ \citenamefont
		{Richter}}]{Russell2016SSR}%
	\BibitemOpen
	\bibfield  {author} {\bibinfo {author} {\bibfnamefont {C.~T.}\ \bibnamefont
			{Russell}}, \bibinfo {author} {\bibfnamefont {B.~J.}\ \bibnamefont
			{Anderson}}, \bibinfo {author} {\bibfnamefont {W.}~\bibnamefont
			{Baumjohann}}, \bibinfo {author} {\bibfnamefont {K.~R.}\ \bibnamefont
			{Bromund}}, \bibinfo {author} {\bibfnamefont {D.}~\bibnamefont {Dearborn}},
		\bibinfo {author} {\bibfnamefont {D.}~\bibnamefont {Fischer}}, \bibinfo
		{author} {\bibfnamefont {G.}~\bibnamefont {Le}}, \bibinfo {author}
		{\bibfnamefont {H.~K.}\ \bibnamefont {Leinweber}}, \bibinfo {author}
		{\bibfnamefont {D.}~\bibnamefont {Leneman}}, \bibinfo {author} {\bibfnamefont
			{W.}~\bibnamefont {Magnes}}, \bibinfo {author} {\bibfnamefont {J.~D.}\
			\bibnamefont {Means}}, \bibinfo {author} {\bibfnamefont {M.~B.}\ \bibnamefont
			{Moldwin}}, \bibinfo {author} {\bibfnamefont {R.}~\bibnamefont {Nakamura}},
		\bibinfo {author} {\bibfnamefont {D.}~\bibnamefont {Pierce}}, \bibinfo
		{author} {\bibfnamefont {F.}~\bibnamefont {Plaschke}}, \bibinfo {author}
		{\bibfnamefont {K.~M.}\ \bibnamefont {Rowe}}, \bibinfo {author}
		{\bibfnamefont {J.~A.}\ \bibnamefont {Slavin}}, \bibinfo {author}
		{\bibfnamefont {R.~J.}\ \bibnamefont {Strangeway}}, \bibinfo {author}
		{\bibfnamefont {R.}~\bibnamefont {Torbert}}, \bibinfo {author} {\bibfnamefont
			{C.}~\bibnamefont {Hagen}}, \bibinfo {author} {\bibfnamefont
			{I.}~\bibnamefont {Jernej}}, \bibinfo {author} {\bibfnamefont
			{A.}~\bibnamefont {Valavanoglou}}, \ and\ \bibinfo {author} {\bibfnamefont
			{I.}~\bibnamefont {Richter}},\ }\bibfield  {title} {\enquote {\bibinfo
			{title} {The magnetospheric multiscale magnetometers},}\ }\href {\doibase
		10.1007/s11214-014-0057-3} {\bibfield  {journal} {\bibinfo  {journal} {Space
				Science Reviews}\ }\textbf {\bibinfo {volume} {199}},\ \bibinfo {pages}
		{189--256} (\bibinfo {year} {2016})}\BibitemShut {NoStop}%
	\bibitem [{\citenamefont {Torbert}\ \emph {et~al.}(2016)\citenamefont
		{Torbert}, \citenamefont {Russell}, \citenamefont {Magnes}, \citenamefont
		{Ergun}, \citenamefont {Lindqvist}, \citenamefont {LeContel}, \citenamefont
		{Vaith}, \citenamefont {Macri}, \citenamefont {Myers}, \citenamefont {Rau},
		\citenamefont {Needell}, \citenamefont {King}, \citenamefont {Granoff},
		\citenamefont {Chutter}, \citenamefont {Dors}, \citenamefont {Olsson},
		\citenamefont {Khotyaintsev}, \citenamefont {Eriksson}, \citenamefont
		{Kletzing}, \citenamefont {Bounds}, \citenamefont {Anderson}, \citenamefont
		{Baumjohann}, \citenamefont {Steller}, \citenamefont {Bromund}, \citenamefont
		{Le}, \citenamefont {Nakamura}, \citenamefont {Strangeway}, \citenamefont
		{Leinweber}, \citenamefont {Tucker}, \citenamefont {Westfall}, \citenamefont
		{Fischer}, \citenamefont {Plaschke}, \citenamefont {Porter},\ and\
		\citenamefont {Lappalainen}}]{Torbert2016SSR}%
	\BibitemOpen
	\bibfield  {author} {\bibinfo {author} {\bibfnamefont {R.~B.}\ \bibnamefont
			{Torbert}}, \bibinfo {author} {\bibfnamefont {C.~T.}\ \bibnamefont
			{Russell}}, \bibinfo {author} {\bibfnamefont {W.}~\bibnamefont {Magnes}},
		\bibinfo {author} {\bibfnamefont {R.~E.}\ \bibnamefont {Ergun}}, \bibinfo
		{author} {\bibfnamefont {P.-A.}\ \bibnamefont {Lindqvist}}, \bibinfo {author}
		{\bibfnamefont {O.}~\bibnamefont {LeContel}}, \bibinfo {author}
		{\bibfnamefont {H.}~\bibnamefont {Vaith}}, \bibinfo {author} {\bibfnamefont
			{J.}~\bibnamefont {Macri}}, \bibinfo {author} {\bibfnamefont
			{S.}~\bibnamefont {Myers}}, \bibinfo {author} {\bibfnamefont
			{D.}~\bibnamefont {Rau}}, \bibinfo {author} {\bibfnamefont {J.}~\bibnamefont
			{Needell}}, \bibinfo {author} {\bibfnamefont {B.}~\bibnamefont {King}},
		\bibinfo {author} {\bibfnamefont {M.}~\bibnamefont {Granoff}}, \bibinfo
		{author} {\bibfnamefont {M.}~\bibnamefont {Chutter}}, \bibinfo {author}
		{\bibfnamefont {I.}~\bibnamefont {Dors}}, \bibinfo {author} {\bibfnamefont
			{G.}~\bibnamefont {Olsson}}, \bibinfo {author} {\bibfnamefont {Y.~V.}\
			\bibnamefont {Khotyaintsev}}, \bibinfo {author} {\bibfnamefont
			{A.}~\bibnamefont {Eriksson}}, \bibinfo {author} {\bibfnamefont {C.~A.}\
			\bibnamefont {Kletzing}}, \bibinfo {author} {\bibfnamefont {S.}~\bibnamefont
			{Bounds}}, \bibinfo {author} {\bibfnamefont {B.}~\bibnamefont {Anderson}},
		\bibinfo {author} {\bibfnamefont {W.}~\bibnamefont {Baumjohann}}, \bibinfo
		{author} {\bibfnamefont {M.}~\bibnamefont {Steller}}, \bibinfo {author}
		{\bibfnamefont {K.}~\bibnamefont {Bromund}}, \bibinfo {author} {\bibfnamefont
			{G.}~\bibnamefont {Le}}, \bibinfo {author} {\bibfnamefont {R.}~\bibnamefont
			{Nakamura}}, \bibinfo {author} {\bibfnamefont {R.~J.}\ \bibnamefont
			{Strangeway}}, \bibinfo {author} {\bibfnamefont {H.~K.}\ \bibnamefont
			{Leinweber}}, \bibinfo {author} {\bibfnamefont {S.}~\bibnamefont {Tucker}},
		\bibinfo {author} {\bibfnamefont {J.}~\bibnamefont {Westfall}}, \bibinfo
		{author} {\bibfnamefont {D.}~\bibnamefont {Fischer}}, \bibinfo {author}
		{\bibfnamefont {F.}~\bibnamefont {Plaschke}}, \bibinfo {author}
		{\bibfnamefont {J.}~\bibnamefont {Porter}}, \ and\ \bibinfo {author}
		{\bibfnamefont {K.}~\bibnamefont {Lappalainen}},\ }\bibfield  {title}
	{\enquote {\bibinfo {title} {The fields instrument suite on mms: Scientific
				objectives, measurements, and data products},}\ }\href {\doibase
		10.1007/s11214-014-0109-8} {\bibfield  {journal} {\bibinfo  {journal} {Space
				Science Reviews}\ }\textbf {\bibinfo {volume} {199}},\ \bibinfo {pages}
		{105--135} (\bibinfo {year} {2016})}\BibitemShut {NoStop}%
	\bibitem [{\citenamefont {{Farris}}\ and\ \citenamefont
		{{Russell}}(1994)}]{Farris1994JGR}%
	\BibitemOpen
	\bibfield  {author} {\bibinfo {author} {\bibfnamefont {M.~H.}\ \bibnamefont
			{{Farris}}}\ and\ \bibinfo {author} {\bibfnamefont {C.~T.}\ \bibnamefont
			{{Russell}}},\ }\bibfield  {title} {\enquote {\bibinfo {title} {{Determining
					the standoff distance of the bow shock: Mach number dependence and use of
					models}},}\ }\href {\doibase 10.1029/94JA01020} {\bibfield  {journal}
		{\bibinfo  {journal} {Journal of Geophysical Research: Space Physics}\
		}\textbf {\bibinfo {volume} {99}},\ \bibinfo {pages} {17} (\bibinfo {year}
		{1994})}\BibitemShut {NoStop}%
	\bibitem [{\citenamefont {{Shue}}\ \emph {et~al.}(1998)\citenamefont {{Shue}},
		\citenamefont {{Song}}, \citenamefont {{Russell}}, \citenamefont
		{{Steinberg}}, \citenamefont {{Chao}}, \citenamefont {{Zastenker}},
		\citenamefont {{Vaisberg}}, \citenamefont {{Kokubun}}, \citenamefont
		{{Singer}}, \citenamefont {{Detman}},\ and\ \citenamefont
		{{Kawano}}}]{shue1998JGR}%
	\BibitemOpen
	\bibfield  {author} {\bibinfo {author} {\bibfnamefont {J.-H.}\ \bibnamefont
			{{Shue}}}, \bibinfo {author} {\bibfnamefont {P.}~\bibnamefont {{Song}}},
		\bibinfo {author} {\bibfnamefont {C.~T.}\ \bibnamefont {{Russell}}}, \bibinfo
		{author} {\bibfnamefont {J.~T.}\ \bibnamefont {{Steinberg}}}, \bibinfo
		{author} {\bibfnamefont {J.~K.}\ \bibnamefont {{Chao}}}, \bibinfo {author}
		{\bibfnamefont {G.}~\bibnamefont {{Zastenker}}}, \bibinfo {author}
		{\bibfnamefont {O.~L.}\ \bibnamefont {{Vaisberg}}}, \bibinfo {author}
		{\bibfnamefont {S.}~\bibnamefont {{Kokubun}}}, \bibinfo {author}
		{\bibfnamefont {H.~J.}\ \bibnamefont {{Singer}}}, \bibinfo {author}
		{\bibfnamefont {T.~R.}\ \bibnamefont {{Detman}}}, \ and\ \bibinfo {author}
		{\bibfnamefont {H.}~\bibnamefont {{Kawano}}},\ }\bibfield  {title} {\enquote
		{\bibinfo {title} {{Magnetopause location under extreme solar wind
					conditions}},}\ }\href {\doibase 10.1029/98JA01103} {\bibfield  {journal}
		{\bibinfo  {journal} {Journal of Geophysical Research: Space Physics}\
		}\textbf {\bibinfo {volume} {103}},\ \bibinfo {pages} {17691--17700}
		(\bibinfo {year} {1998})}\BibitemShut {NoStop}%
	\bibitem [{\citenamefont {{Riedl}}, \citenamefont {{M{\"u}ller}},\ and\
		\citenamefont {{Wessel}}(2013)}]{Riedl2013EPJST_entropy}%
	\BibitemOpen
	\bibfield  {author} {\bibinfo {author} {\bibfnamefont {M.}~\bibnamefont
			{{Riedl}}}, \bibinfo {author} {\bibfnamefont {A.}~\bibnamefont
			{{M{\"u}ller}}}, \ and\ \bibinfo {author} {\bibfnamefont {N.}~\bibnamefont
			{{Wessel}}},\ }\bibfield  {title} {\enquote {\bibinfo {title} {{Practical
					considerations of permutation entropy. A tutorial review}},}\ }\href
	{\doibase 10.1140/epjst/e2013-01862-7} {\bibfield  {journal} {\bibinfo
			{journal} {European Physical Journal Special Topics}\ }\textbf {\bibinfo
			{volume} {222}},\ \bibinfo {pages} {249--262} (\bibinfo {year}
		{2013})}\BibitemShut {NoStop}%
	\bibitem [{\citenamefont {Osmane}, \citenamefont {Dimmock},\ and\ \citenamefont
		{Pulkkinen}(2019)}]{Osmane2019JGR_complexity}%
	\BibitemOpen
	\bibfield  {author} {\bibinfo {author} {\bibfnamefont {A.}~\bibnamefont
			{Osmane}}, \bibinfo {author} {\bibfnamefont {A.~P.}\ \bibnamefont {Dimmock}},
		\ and\ \bibinfo {author} {\bibfnamefont {T.~I.}\ \bibnamefont {Pulkkinen}},\
	}\bibfield  {title} {\enquote {\bibinfo {title} {Jensen-shannon complexity
				and permutation entropy analysis of geomagnetic auroral currents},}\ }\href
	{\doibase https://doi.org/10.1029/2018JA026248} {\bibfield  {journal}
		{\bibinfo  {journal} {Journal of Geophysical Research: Space Physics}\
		}\textbf {\bibinfo {volume} {124}},\ \bibinfo {pages} {2541--2551} (\bibinfo
		{year} {2019})}\BibitemShut {NoStop}%
	\bibitem [{\citenamefont {Good}\ \emph {et~al.}(2020)\citenamefont {Good},
		\citenamefont {Ala-Lahti}, \citenamefont {Palmerio}, \citenamefont {Kilpua},\
		and\ \citenamefont {Osmane}}]{Good2020ApJ_ICME}%
	\BibitemOpen
	\bibfield  {author} {\bibinfo {author} {\bibfnamefont {S.~W.}\ \bibnamefont
			{Good}}, \bibinfo {author} {\bibfnamefont {M.}~\bibnamefont {Ala-Lahti}},
		\bibinfo {author} {\bibfnamefont {E.}~\bibnamefont {Palmerio}}, \bibinfo
		{author} {\bibfnamefont {E.~K.~J.}\ \bibnamefont {Kilpua}}, \ and\ \bibinfo
		{author} {\bibfnamefont {A.}~\bibnamefont {Osmane}},\ }\bibfield  {title}
	{\enquote {\bibinfo {title} {Radial evolution of magnetic field fluctuations
				in an interplanetary coronal mass ejection sheath},}\ }\href {\doibase
		10.3847/1538-4357/ab7fa2} {\bibfield  {journal} {\bibinfo  {journal} {The
				Astrophysical Journal}\ }\textbf {\bibinfo {volume} {893}},\ \bibinfo {pages}
		{110} (\bibinfo {year} {2020})}\BibitemShut {NoStop}%
	\bibitem [{\citenamefont {Miranda}\ \emph {et~al.}(2021)\citenamefont
		{Miranda}, \citenamefont {Valdivia}, \citenamefont {Chian},\ and\
		\citenamefont {Muñoz}}]{Miranda2021ApJ_complexity}%
	\BibitemOpen
	\bibfield  {author} {\bibinfo {author} {\bibfnamefont {R.~A.}\ \bibnamefont
			{Miranda}}, \bibinfo {author} {\bibfnamefont {J.~A.}\ \bibnamefont
			{Valdivia}}, \bibinfo {author} {\bibfnamefont {A.~C.-L.}\ \bibnamefont
			{Chian}}, \ and\ \bibinfo {author} {\bibfnamefont {P.~R.}\ \bibnamefont
			{Muñoz}},\ }\bibfield  {title} {\enquote {\bibinfo {title} {Complexity of
				magnetic-field turbulence at reconnection exhausts in the solar wind at 1
				au},}\ }\href {\doibase 10.3847/1538-4357/ac2dfe} {\bibfield  {journal}
		{\bibinfo  {journal} {The Astrophysical Journal}\ }\textbf {\bibinfo {volume}
			{923}},\ \bibinfo {pages} {132} (\bibinfo {year} {2021})}\BibitemShut
	{NoStop}%
	\bibitem [{\citenamefont {{Kilpua, E. K. J.}}\ \emph
		{et~al.}(2022)\citenamefont {{Kilpua, E. K. J.}}, \citenamefont {{Good, S.
				W.}}, \citenamefont {{Ala-Lahti, M.}}, \citenamefont {{Osmane, A.}},
		\citenamefont {{Pal, S.}}, \citenamefont {{Soljento, J. E.}}, \citenamefont
		{{Zhao, L. L.}},\ and\ \citenamefont {{Bale, S.}}}]{Kilpua2022_ICME}%
	\BibitemOpen
	\bibfield  {author} {\bibinfo {author} {\bibnamefont {{Kilpua, E. K. J.}}},
		\bibinfo {author} {\bibnamefont {{Good, S. W.}}}, \bibinfo {author}
		{\bibnamefont {{Ala-Lahti, M.}}}, \bibinfo {author} {\bibnamefont {{Osmane,
					A.}}}, \bibinfo {author} {\bibnamefont {{Pal, S.}}}, \bibinfo {author}
		{\bibnamefont {{Soljento, J. E.}}}, \bibinfo {author} {\bibnamefont {{Zhao,
					L. L.}}}, \ and\ \bibinfo {author} {\bibnamefont {{Bale, S.}}},\ }\bibfield
	{title} {\enquote {\bibinfo {title} {Structure and fluctuations of a slow
				icme sheath observed at 0.5 au by the parker solar probe},}\ }\href {\doibase
		10.1051/0004-6361/202142191} {\bibfield  {journal} {\bibinfo  {journal}
			{Astronomy and Astrophysics}\ }\textbf {\bibinfo {volume} {663}},\ \bibinfo
		{pages} {A108} (\bibinfo {year} {2022})}\BibitemShut {NoStop}%
	\bibitem [{\citenamefont {Kiyani}, \citenamefont {Osman},\ and\ \citenamefont
		{Chapman}(2015)}]{Kiyani2015PTRSA}%
	\BibitemOpen
	\bibfield  {author} {\bibinfo {author} {\bibfnamefont {K.~H.}\ \bibnamefont
			{Kiyani}}, \bibinfo {author} {\bibfnamefont {K.~T.}\ \bibnamefont {Osman}}, \
		and\ \bibinfo {author} {\bibfnamefont {S.~C.}\ \bibnamefont {Chapman}},\
	}\bibfield  {title} {\enquote {\bibinfo {title} {Dissipation and heating in
				solar wind turbulence: from the macro to the micro and back again},}\ }\href
	{\doibase 10.1098/rsta.2014.0155} {\bibfield  {journal} {\bibinfo  {journal}
			{Philosophical Transactions of the Royal Society A: Mathematical, Physical
				and Engineering Sciences}\ }\textbf {\bibinfo {volume} {373}},\ \bibinfo
		{pages} {20140155} (\bibinfo {year} {2015})}\BibitemShut {NoStop}%
	\bibitem [{\citenamefont {Maruca}\ \emph {et~al.}(2018)\citenamefont {Maruca},
		\citenamefont {Chasapis}, \citenamefont {Gary}, \citenamefont
		{Bandyopadhyay}, \citenamefont {Chhiber}, \citenamefont {Parashar},
		\citenamefont {Matthaeus}, \citenamefont {Shay}, \citenamefont {Burch},
		\citenamefont {Moore}, \citenamefont {Pollock}, \citenamefont {Giles},
		\citenamefont {Paterson}, \citenamefont {Dorelli}, \citenamefont {Gershman},
		\citenamefont {Torbert}, \citenamefont {Russell},\ and\ \citenamefont
		{Strangeway}}]{Maruca2018ApJ}%
	\BibitemOpen
	\bibfield  {author} {\bibinfo {author} {\bibfnamefont {B.~A.}\ \bibnamefont
			{Maruca}}, \bibinfo {author} {\bibfnamefont {A.}~\bibnamefont {Chasapis}},
		\bibinfo {author} {\bibfnamefont {S.~P.}\ \bibnamefont {Gary}}, \bibinfo
		{author} {\bibfnamefont {R.}~\bibnamefont {Bandyopadhyay}}, \bibinfo {author}
		{\bibfnamefont {R.}~\bibnamefont {Chhiber}}, \bibinfo {author} {\bibfnamefont
			{T.~N.}\ \bibnamefont {Parashar}}, \bibinfo {author} {\bibfnamefont {W.~H.}\
			\bibnamefont {Matthaeus}}, \bibinfo {author} {\bibfnamefont {M.~A.}\
			\bibnamefont {Shay}}, \bibinfo {author} {\bibfnamefont {J.~L.}\ \bibnamefont
			{Burch}}, \bibinfo {author} {\bibfnamefont {T.~E.}\ \bibnamefont {Moore}},
		\bibinfo {author} {\bibfnamefont {C.~J.}\ \bibnamefont {Pollock}}, \bibinfo
		{author} {\bibfnamefont {B.~J.}\ \bibnamefont {Giles}}, \bibinfo {author}
		{\bibfnamefont {W.~R.}\ \bibnamefont {Paterson}}, \bibinfo {author}
		{\bibfnamefont {J.}~\bibnamefont {Dorelli}}, \bibinfo {author} {\bibfnamefont
			{D.~J.}\ \bibnamefont {Gershman}}, \bibinfo {author} {\bibfnamefont {R.~B.}\
			\bibnamefont {Torbert}}, \bibinfo {author} {\bibfnamefont {C.~T.}\
			\bibnamefont {Russell}}, \ and\ \bibinfo {author} {\bibfnamefont {R.~J.}\
			\bibnamefont {Strangeway}},\ }\bibfield  {title} {\enquote {\bibinfo {title}
			{Mms observations of beta-dependent contraints on ion temperature-anisotropy
				in earth's magnetosheath},}\ }\href {\doibase 10.3847/1538-4357/aaddfb}
	{\bibfield  {journal} {\bibinfo  {journal} {The Astrophysical Journal}\
		}\textbf {\bibinfo {volume} {866}},\ \bibinfo {pages} {25} (\bibinfo {year}
		{2018})}\BibitemShut {NoStop}%
	\bibitem [{\citenamefont {Huang}\ \emph {et~al.}(2021)\citenamefont {Huang},
		\citenamefont {Xiong}, \citenamefont {Yuan}, \citenamefont {Zhan},
		\citenamefont {Deng}, \citenamefont {Jiang}, \citenamefont {Zhao},
		\citenamefont {Zhang}, \citenamefont {Xu}, \citenamefont {Wei}, \citenamefont
		{Yu},\ and\ \citenamefont {Lin}}]{Huang2021JGR_anisotropy}%
	\BibitemOpen
	\bibfield  {author} {\bibinfo {author} {\bibfnamefont {S.~Y.}\ \bibnamefont
			{Huang}}, \bibinfo {author} {\bibfnamefont {Q.~Y.}\ \bibnamefont {Xiong}},
		\bibinfo {author} {\bibfnamefont {Z.~G.}\ \bibnamefont {Yuan}}, \bibinfo
		{author} {\bibfnamefont {H.~L.}\ \bibnamefont {Zhan}}, \bibinfo {author}
		{\bibfnamefont {X.~H.}\ \bibnamefont {Deng}}, \bibinfo {author}
		{\bibfnamefont {K.}~\bibnamefont {Jiang}}, \bibinfo {author} {\bibfnamefont
			{P.~F.}\ \bibnamefont {Zhao}}, \bibinfo {author} {\bibfnamefont
			{J.}~\bibnamefont {Zhang}}, \bibinfo {author} {\bibfnamefont {S.~B.}\
			\bibnamefont {Xu}}, \bibinfo {author} {\bibfnamefont {Y.~Y.}\ \bibnamefont
			{Wei}}, \bibinfo {author} {\bibfnamefont {L.}~\bibnamefont {Yu}}, \ and\
		\bibinfo {author} {\bibfnamefont {T.~R.}\ \bibnamefont {Lin}},\ }\bibfield
	{title} {\enquote {\bibinfo {title} {Multi-spacecraft measurement of
				anisotropic spatial correlation functions at kinetic range in the
				magnetosheath turbulence},}\ }\href {\doibase
		https://doi.org/10.1029/2020JA028780} {\bibfield  {journal} {\bibinfo
			{journal} {Journal of Geophysical Research: Space Physics}\ }\textbf
		{\bibinfo {volume} {126}},\ \bibinfo {pages} {e2020JA028780} (\bibinfo {year}
		{2021})}\BibitemShut {NoStop}%
	\bibitem [{\citenamefont {Osmane}, \citenamefont {Dimmock},\ and\ \citenamefont
		{Pulkkinen}(2015)}]{Osmane2015GRL_mirror-mode}%
	\BibitemOpen
	\bibfield  {author} {\bibinfo {author} {\bibfnamefont {A.}~\bibnamefont
			{Osmane}}, \bibinfo {author} {\bibfnamefont {A.~P.}\ \bibnamefont {Dimmock}},
		\ and\ \bibinfo {author} {\bibfnamefont {T.~I.}\ \bibnamefont {Pulkkinen}},\
	}\bibfield  {title} {\enquote {\bibinfo {title} {Universal properties of
				mirror mode turbulence in the earth's magnetosheath},}\ }\href {\doibase
		https://doi.org/10.1002/2015GL063771} {\bibfield  {journal} {\bibinfo
			{journal} {Geophysical Research Letters}\ }\textbf {\bibinfo {volume} {42}},\
		\bibinfo {pages} {3085--3092} (\bibinfo {year} {2015})}\BibitemShut {NoStop}%
	\bibitem [{\citenamefont {Lepping}\ \emph {et~al.}(1995)\citenamefont
		{Lepping}, \citenamefont {Ac{\~{u}}na}, \citenamefont {Burlaga},
		\citenamefont {Farrell}, \citenamefont {Slavin}, \citenamefont {Schatten},
		\citenamefont {Mariani}, \citenamefont {Ness}, \citenamefont {Neubauer},
		\citenamefont {Whang}, \citenamefont {Byrnes}, \citenamefont {Kennon},
		\citenamefont {Panetta}, \citenamefont {Scheifele},\ and\ \citenamefont
		{Worley}}]{Lepping1995SSR}%
	\BibitemOpen
	\bibfield  {author} {\bibinfo {author} {\bibfnamefont {R.~P.}\ \bibnamefont
			{Lepping}}, \bibinfo {author} {\bibfnamefont {M.~H.}\ \bibnamefont
			{Ac{\~{u}}na}}, \bibinfo {author} {\bibfnamefont {L.~F.}\ \bibnamefont
			{Burlaga}}, \bibinfo {author} {\bibfnamefont {W.~M.}\ \bibnamefont
			{Farrell}}, \bibinfo {author} {\bibfnamefont {J.~A.}\ \bibnamefont {Slavin}},
		\bibinfo {author} {\bibfnamefont {K.~H.}\ \bibnamefont {Schatten}}, \bibinfo
		{author} {\bibfnamefont {F.}~\bibnamefont {Mariani}}, \bibinfo {author}
		{\bibfnamefont {N.~F.}\ \bibnamefont {Ness}}, \bibinfo {author}
		{\bibfnamefont {F.~M.}\ \bibnamefont {Neubauer}}, \bibinfo {author}
		{\bibfnamefont {Y.~C.}\ \bibnamefont {Whang}}, \bibinfo {author}
		{\bibfnamefont {J.~B.}\ \bibnamefont {Byrnes}}, \bibinfo {author}
		{\bibfnamefont {R.~S.}\ \bibnamefont {Kennon}}, \bibinfo {author}
		{\bibfnamefont {P.~V.}\ \bibnamefont {Panetta}}, \bibinfo {author}
		{\bibfnamefont {J.}~\bibnamefont {Scheifele}}, \ and\ \bibinfo {author}
		{\bibfnamefont {E.~M.}\ \bibnamefont {Worley}},\ }\bibfield  {title}
	{\enquote {\bibinfo {title} {The wind magnetic field investigation},}\ }\href
	{\doibase 10.1007/BF00751330} {\bibfield  {journal} {\bibinfo  {journal}
			{Space Science Reviews}\ }\textbf {\bibinfo {volume} {71}},\ \bibinfo {pages}
		{207--229} (\bibinfo {year} {1995})}\BibitemShut {NoStop}%
	\bibitem [{\citenamefont {Ogilvie}\ \emph {et~al.}(1995)\citenamefont
		{Ogilvie}, \citenamefont {Chornay}, \citenamefont {Fritzenreiter},
		\citenamefont {Hunsaker}, \citenamefont {Keller}, \citenamefont {Lobell},
		\citenamefont {Miller}, \citenamefont {Scudder}, \citenamefont {Sittler},
		\citenamefont {Torbert}, \citenamefont {Bodet}, \citenamefont {Needell},
		\citenamefont {Lazarus}, \citenamefont {Steinberg}, \citenamefont {Tappan},
		\citenamefont {Mavretic},\ and\ \citenamefont {Gergin}}]{Ogilvie1995SSR}%
	\BibitemOpen
	\bibfield  {author} {\bibinfo {author} {\bibfnamefont {K.~W.}\ \bibnamefont
			{Ogilvie}}, \bibinfo {author} {\bibfnamefont {D.~J.}\ \bibnamefont
			{Chornay}}, \bibinfo {author} {\bibfnamefont {R.~J.}\ \bibnamefont
			{Fritzenreiter}}, \bibinfo {author} {\bibfnamefont {F.}~\bibnamefont
			{Hunsaker}}, \bibinfo {author} {\bibfnamefont {J.}~\bibnamefont {Keller}},
		\bibinfo {author} {\bibfnamefont {J.}~\bibnamefont {Lobell}}, \bibinfo
		{author} {\bibfnamefont {G.}~\bibnamefont {Miller}}, \bibinfo {author}
		{\bibfnamefont {J.~D.}\ \bibnamefont {Scudder}}, \bibinfo {author}
		{\bibfnamefont {E.~C.}\ \bibnamefont {Sittler}}, \bibinfo {author}
		{\bibfnamefont {R.~B.}\ \bibnamefont {Torbert}}, \bibinfo {author}
		{\bibfnamefont {D.}~\bibnamefont {Bodet}}, \bibinfo {author} {\bibfnamefont
			{G.}~\bibnamefont {Needell}}, \bibinfo {author} {\bibfnamefont {A.~J.}\
			\bibnamefont {Lazarus}}, \bibinfo {author} {\bibfnamefont {J.~T.}\
			\bibnamefont {Steinberg}}, \bibinfo {author} {\bibfnamefont {J.~H.}\
			\bibnamefont {Tappan}}, \bibinfo {author} {\bibfnamefont {A.}~\bibnamefont
			{Mavretic}}, \ and\ \bibinfo {author} {\bibfnamefont {E.}~\bibnamefont
			{Gergin}},\ }\bibfield  {title} {\enquote {\bibinfo {title} {Swe, a
				comprehensive plasma instrument for the wind spacecraft},}\ }\href {\doibase
		10.1007/BF00751326} {\bibfield  {journal} {\bibinfo  {journal} {Space Science
				Reviews}\ }\textbf {\bibinfo {volume} {71}},\ \bibinfo {pages} {55--77}
		(\bibinfo {year} {1995})}\BibitemShut {NoStop}%
	\bibitem [{\citenamefont {Pollock}\ \emph {et~al.}(2016)\citenamefont
		{Pollock}, \citenamefont {Moore}, \citenamefont {Jacques}, \citenamefont
		{Burch}, \citenamefont {Gliese}, \citenamefont {Saito}, \citenamefont
		{Omoto}, \citenamefont {Avanov}, \citenamefont {Barrie}, \citenamefont
		{Coffey}, \citenamefont {Dorelli}, \citenamefont {Gershman}, \citenamefont
		{Giles}, \citenamefont {Rosnack}, \citenamefont {Salo}, \citenamefont
		{Yokota}, \citenamefont {Adrian}, \citenamefont {Aoustin}, \citenamefont
		{Auletti}, \citenamefont {Aung}, \citenamefont {Bigio}, \citenamefont {Cao},
		\citenamefont {Chandler}, \citenamefont {Chornay}, \citenamefont {Christian},
		\citenamefont {Clark}, \citenamefont {Collinson}, \citenamefont {Corris},
		\citenamefont {De Los Santos}, \citenamefont {Devlin}, \citenamefont
		{Diaz}, \citenamefont {Dickerson}, \citenamefont {Dickson}, \citenamefont
		{Diekmann}, \citenamefont {Diggs}, \citenamefont {Duncan}, \citenamefont
		{Figueroa-Vinas}, \citenamefont {Firman}, \citenamefont {Freeman},
		\citenamefont {Galassi}, \citenamefont {Garcia}, \citenamefont {Goodhart},
		\citenamefont {Guererro}, \citenamefont {Hageman}, \citenamefont {Hanley},
		\citenamefont {Hemminger}, \citenamefont {Holland}, \citenamefont {Hutchins},
		\citenamefont {James}, \citenamefont {Jones}, \citenamefont {Kreisler},
		\citenamefont {Kujawski}, \citenamefont {Lavu}, \citenamefont {Lobell},
		\citenamefont {LeCompte}, \citenamefont {Lukemire}, \citenamefont
		{MacDonald}, \citenamefont {Mariano}, \citenamefont {Mukai}, \citenamefont
		{Narayanan}, \citenamefont {Nguyan}, \citenamefont {Onizuka}, \citenamefont
		{Paterson}, \citenamefont {Persyn}, \citenamefont {Piepgrass}, \citenamefont
		{Cheney}, \citenamefont {Rager}, \citenamefont {Raghuram}, \citenamefont
		{Ramil}, \citenamefont {Reichenthal}, \citenamefont {Rodriguez},
		\citenamefont {Rouzaud}, \citenamefont {Rucker}, \citenamefont {Saito},
		\citenamefont {Samara}, \citenamefont {Sauvaud}, \citenamefont {Schuster},
		\citenamefont {Shappirio}, \citenamefont {Shelton}, \citenamefont {Sher},
		\citenamefont {Smith}, \citenamefont {Smith}, \citenamefont {Smith},
		\citenamefont {Steinfeld}, \citenamefont {Szymkiewicz}, \citenamefont
		{Tanimoto}, \citenamefont {Taylor}, \citenamefont {Tucker}, \citenamefont
		{Tull}, \citenamefont {Uhl}, \citenamefont {Vloet}, \citenamefont {Walpole},
		\citenamefont {Weidner}, \citenamefont {White}, \citenamefont {Winkert},
		\citenamefont {Yeh},\ and\ \citenamefont {Zeuch}}]{Pollock2016SSR}%
	\BibitemOpen
	\bibfield  {author} {\bibinfo {author} {\bibfnamefont {C.}~\bibnamefont
			{Pollock}}, \bibinfo {author} {\bibfnamefont {T.}~\bibnamefont {Moore}},
		\bibinfo {author} {\bibfnamefont {A.}~\bibnamefont {Jacques}}, \bibinfo
		{author} {\bibfnamefont {J.}~\bibnamefont {Burch}}, \bibinfo {author}
		{\bibfnamefont {U.}~\bibnamefont {Gliese}}, \bibinfo {author} {\bibfnamefont
			{Y.}~\bibnamefont {Saito}}, \bibinfo {author} {\bibfnamefont
			{T.}~\bibnamefont {Omoto}}, \bibinfo {author} {\bibfnamefont
			{L.}~\bibnamefont {Avanov}}, \bibinfo {author} {\bibfnamefont
			{A.}~\bibnamefont {Barrie}}, \bibinfo {author} {\bibfnamefont
			{V.}~\bibnamefont {Coffey}}, \bibinfo {author} {\bibfnamefont
			{J.}~\bibnamefont {Dorelli}}, \bibinfo {author} {\bibfnamefont
			{D.}~\bibnamefont {Gershman}}, \bibinfo {author} {\bibfnamefont
			{B.}~\bibnamefont {Giles}}, \bibinfo {author} {\bibfnamefont
			{T.}~\bibnamefont {Rosnack}}, \bibinfo {author} {\bibfnamefont
			{C.}~\bibnamefont {Salo}}, \bibinfo {author} {\bibfnamefont {S.}~\bibnamefont
			{Yokota}}, \bibinfo {author} {\bibfnamefont {M.}~\bibnamefont {Adrian}},
		\bibinfo {author} {\bibfnamefont {C.}~\bibnamefont {Aoustin}}, \bibinfo
		{author} {\bibfnamefont {C.}~\bibnamefont {Auletti}}, \bibinfo {author}
		{\bibfnamefont {S.}~\bibnamefont {Aung}}, \bibinfo {author} {\bibfnamefont
			{V.}~\bibnamefont {Bigio}}, \bibinfo {author} {\bibfnamefont
			{N.}~\bibnamefont {Cao}}, \bibinfo {author} {\bibfnamefont {M.}~\bibnamefont
			{Chandler}}, \bibinfo {author} {\bibfnamefont {D.}~\bibnamefont {Chornay}},
		\bibinfo {author} {\bibfnamefont {K.}~\bibnamefont {Christian}}, \bibinfo
		{author} {\bibfnamefont {G.}~\bibnamefont {Clark}}, \bibinfo {author}
		{\bibfnamefont {G.}~\bibnamefont {Collinson}}, \bibinfo {author}
		{\bibfnamefont {T.}~\bibnamefont {Corris}}, \bibinfo {author} {\bibfnamefont
			{A.}~\bibnamefont {De Los Santos}}, \bibinfo {author} {\bibfnamefont
			{R.}~\bibnamefont {Devlin}}, \bibinfo {author} {\bibfnamefont
			{T.}~\bibnamefont {Diaz}}, \bibinfo {author} {\bibfnamefont {T.}~\bibnamefont
			{Dickerson}}, \bibinfo {author} {\bibfnamefont {C.}~\bibnamefont {Dickson}},
		\bibinfo {author} {\bibfnamefont {A.}~\bibnamefont {Diekmann}}, \bibinfo
		{author} {\bibfnamefont {F.}~\bibnamefont {Diggs}}, \bibinfo {author}
		{\bibfnamefont {C.}~\bibnamefont {Duncan}}, \bibinfo {author} {\bibfnamefont
			{A.}~\bibnamefont {Figueroa-Vinas}}, \bibinfo {author} {\bibfnamefont
			{C.}~\bibnamefont {Firman}}, \bibinfo {author} {\bibfnamefont
			{M.}~\bibnamefont {Freeman}}, \bibinfo {author} {\bibfnamefont
			{N.}~\bibnamefont {Galassi}}, \bibinfo {author} {\bibfnamefont
			{K.}~\bibnamefont {Garcia}}, \bibinfo {author} {\bibfnamefont
			{G.}~\bibnamefont {Goodhart}}, \bibinfo {author} {\bibfnamefont
			{D.}~\bibnamefont {Guererro}}, \bibinfo {author} {\bibfnamefont
			{J.}~\bibnamefont {Hageman}}, \bibinfo {author} {\bibfnamefont
			{J.}~\bibnamefont {Hanley}}, \bibinfo {author} {\bibfnamefont
			{E.}~\bibnamefont {Hemminger}}, \bibinfo {author} {\bibfnamefont
			{M.}~\bibnamefont {Holland}}, \bibinfo {author} {\bibfnamefont
			{M.}~\bibnamefont {Hutchins}}, \bibinfo {author} {\bibfnamefont
			{T.}~\bibnamefont {James}}, \bibinfo {author} {\bibfnamefont
			{W.}~\bibnamefont {Jones}}, \bibinfo {author} {\bibfnamefont
			{S.}~\bibnamefont {Kreisler}}, \bibinfo {author} {\bibfnamefont
			{J.}~\bibnamefont {Kujawski}}, \bibinfo {author} {\bibfnamefont
			{V.}~\bibnamefont {Lavu}}, \bibinfo {author} {\bibfnamefont {J.}~\bibnamefont
			{Lobell}}, \bibinfo {author} {\bibfnamefont {E.}~\bibnamefont {LeCompte}},
		\bibinfo {author} {\bibfnamefont {A.}~\bibnamefont {Lukemire}}, \bibinfo
		{author} {\bibfnamefont {E.}~\bibnamefont {MacDonald}}, \bibinfo {author}
		{\bibfnamefont {A.}~\bibnamefont {Mariano}}, \bibinfo {author} {\bibfnamefont
			{T.}~\bibnamefont {Mukai}}, \bibinfo {author} {\bibfnamefont
			{K.}~\bibnamefont {Narayanan}}, \bibinfo {author} {\bibfnamefont
			{Q.}~\bibnamefont {Nguyan}}, \bibinfo {author} {\bibfnamefont
			{M.}~\bibnamefont {Onizuka}}, \bibinfo {author} {\bibfnamefont
			{W.}~\bibnamefont {Paterson}}, \bibinfo {author} {\bibfnamefont
			{S.}~\bibnamefont {Persyn}}, \bibinfo {author} {\bibfnamefont
			{B.}~\bibnamefont {Piepgrass}}, \bibinfo {author} {\bibfnamefont
			{F.}~\bibnamefont {Cheney}}, \bibinfo {author} {\bibfnamefont
			{A.}~\bibnamefont {Rager}}, \bibinfo {author} {\bibfnamefont
			{T.}~\bibnamefont {Raghuram}}, \bibinfo {author} {\bibfnamefont
			{A.}~\bibnamefont {Ramil}}, \bibinfo {author} {\bibfnamefont
			{L.}~\bibnamefont {Reichenthal}}, \bibinfo {author} {\bibfnamefont
			{H.}~\bibnamefont {Rodriguez}}, \bibinfo {author} {\bibfnamefont
			{J.}~\bibnamefont {Rouzaud}}, \bibinfo {author} {\bibfnamefont
			{A.}~\bibnamefont {Rucker}}, \bibinfo {author} {\bibfnamefont
			{Y.}~\bibnamefont {Saito}}, \bibinfo {author} {\bibfnamefont
			{M.}~\bibnamefont {Samara}}, \bibinfo {author} {\bibfnamefont {J.-A.}\
			\bibnamefont {Sauvaud}}, \bibinfo {author} {\bibfnamefont {D.}~\bibnamefont
			{Schuster}}, \bibinfo {author} {\bibfnamefont {M.}~\bibnamefont {Shappirio}},
		\bibinfo {author} {\bibfnamefont {K.}~\bibnamefont {Shelton}}, \bibinfo
		{author} {\bibfnamefont {D.}~\bibnamefont {Sher}}, \bibinfo {author}
		{\bibfnamefont {D.}~\bibnamefont {Smith}}, \bibinfo {author} {\bibfnamefont
			{K.}~\bibnamefont {Smith}}, \bibinfo {author} {\bibfnamefont
			{S.}~\bibnamefont {Smith}}, \bibinfo {author} {\bibfnamefont
			{D.}~\bibnamefont {Steinfeld}}, \bibinfo {author} {\bibfnamefont
			{R.}~\bibnamefont {Szymkiewicz}}, \bibinfo {author} {\bibfnamefont
			{K.}~\bibnamefont {Tanimoto}}, \bibinfo {author} {\bibfnamefont
			{J.}~\bibnamefont {Taylor}}, \bibinfo {author} {\bibfnamefont
			{C.}~\bibnamefont {Tucker}}, \bibinfo {author} {\bibfnamefont
			{K.}~\bibnamefont {Tull}}, \bibinfo {author} {\bibfnamefont {A.}~\bibnamefont
			{Uhl}}, \bibinfo {author} {\bibfnamefont {J.}~\bibnamefont {Vloet}}, \bibinfo
		{author} {\bibfnamefont {P.}~\bibnamefont {Walpole}}, \bibinfo {author}
		{\bibfnamefont {S.}~\bibnamefont {Weidner}}, \bibinfo {author} {\bibfnamefont
			{D.}~\bibnamefont {White}}, \bibinfo {author} {\bibfnamefont
			{G.}~\bibnamefont {Winkert}}, \bibinfo {author} {\bibfnamefont {P.-S.}\
			\bibnamefont {Yeh}}, \ and\ \bibinfo {author} {\bibfnamefont
			{M.}~\bibnamefont {Zeuch}},\ }\bibfield  {title} {\enquote {\bibinfo {title}
			{Fast plasma investigation for magnetospheric multiscale},}\ }\href {\doibase
		10.1007/s11214-016-0245-4} {\bibfield  {journal} {\bibinfo  {journal} {Space
				Science Reviews}\ }\textbf {\bibinfo {volume} {199}},\ \bibinfo {pages}
		{331--406} (\bibinfo {year} {2016})}\BibitemShut {NoStop}%
\end{thebibliography}


\end{document}